\documentclass{PoS}

\usepackage[english]{babel}
\usepackage{amsmath}

\usepackage[comma,square,numbers,sort&compress]{natbib}

\title{(Anti-)nuclei production and flow in pp, p--Pb and Pb--Pb collisions with ALICE}

\ShortTitle{(Anti-)nuclei production and flow with ALICE}

\author{\speaker{Sebastian Hornung} on behalf of the ALICE Collaboration\thanks{Supported by the German Federal Ministry of Education and Research and HGS-HIRe for FAIR}\\
        Physikalisches Institut, Ruprecht-Karls-Universit\"at Heidelberg, Heidelberg, Germany\\
GSI Helmholtz Centre for Heavy Ion Research, Planckstra\ss{}e 1, 64291 Darmstadt, Germany\\
        E-mail: \email{Sebastian.Hornung@cern.ch}}

\abstract{
High energy pp, p--Pb, and Pb--Pb collisions at the LHC offer a unique opportunity to study the production of light (anti-)nuclei. The study of the production yield of (anti-)nuclei in heavy-ion collisions at LHC energies probes the late stages in the evolution of the hot, dense nuclear matter created in the collision.  
Measurements performed in smaller collision systems are crucial to understand how the production mechanism evolves going from small to large systems.
The results on the determination of the (anti-)nuclei yields will be complemented by the measurement of their azimuthal anisotropic production. 
This is a powerful tool to gain insight into the production mechanism of light nuclei in  relativistic ion collisions: in particular, it will help to distinguish between coalescence and hydrodynamic models.
The coalescence parameter and the nuclei-to-proton ratio is studied as a function of the system size, showing a smooth transition from low to high values of the charged-particle multiplicity density. The experimental results can be described with the canonical statistical hadronization model as well as the coalescence approach within uncertainties.
The theoretical description of the new results on the measurement of the elliptic and the triangular flow of deuterons and $^3$He produced in Pb--Pb collisions at $\sqrt{s_\mathrm{NN}} = 5.02$ TeV requires more sophisticated coalescence and statistical hadronization models.
}

\FullConference{
European Physical Society Conference on High Energy Physics - EPS-HEP2019 -\\
			10-17 July, 2019\\
			Ghent, Belgium}

\begin{document}

\section{Production of light (anti-)nuclei in ion collisions}

At the LHC, ligh (anti-)nuclei are produced in measurable amounts in pp, p--Pb and Pb--Pb collisions at different energies.
The production mechanism is one of the open questions in the field of heavy-ion physics because the separation energies of the produced nuclei (a few MeV) are low compared to the temperature of the system ($T_\mathrm{kin} \approx 100$ MeV) in which they are formed. Thus, studying the production and flow of (anti-)nuclei offers a crucial insight into the late stage of the collisions, providing especially information about the process of hadronization, and helps to understand how these objects survive in such extreme conditions.

To describe the measured production yields, two classes of models are used, the statistical hadronization and the coalescence models.
In the Statistical Hadronization Model (SHM) \cite{Andronic:2017pug, BraunMunzinger:2003zd}, the particles are produced from a medium in statistical equilibrium at the chemical freeze-out, when the rate of inelastic collisions becomes negligible.
The yields of hadrons in central \mbox{Pb--Pb} collisions are reproduced by this approach, within uncertainties, using the grand canonical ensemble description \cite{BraunMunzinger:2003zd}. The total production yields have an exponential dependence on the mass of the produced hadron divided by the chemical freeze-out temperature. This leads to a strong sensitivity to $T_\mathrm{chem}$ of the (anti-)nuclei yields. Elastic and quasi-elastic scattering might still occur among hadrons during the further evolution of the system. Thus, the transverse momentum distributions can be modified until also the elastic interactions cease at the kinetic freeze-out.
In smaller systems, like pp and p--Pb collisions, the exact conservation of quantum numbers, such as the baryon number, across the correlation volume $V_{\mathrm{c}}$ has to be taken into account. This is achieved by switching to the canonical ensemble description, which leads to a suppression of the yields for smaller system sizes.

On the other hand, the production of light (anti-)nuclei can be explained via the coalescence of protons and neutrons which are close by in phase space at the kinetic freeze-out and thus form a nucleus \cite{Kapusta:1980zz,Scheibl:1998tk}. The key parameter of these models is the coalescence parameter, which is given by 
\begin{align}
	B_A(\vec{p}_{\mathrm{p}}) = E_\mathrm{i}\frac{\mathrm{d}^{3}N_\mathrm{i}}{\mathrm{d}p_\mathrm{i}^{3}} \Big/ \left(E_{\mathrm{p}}\frac{\mathrm{d}^{3}N_{\mathrm{p}}}{\mathrm{d}p_{\mathrm{p}}^{3}}\right)^{A} \Bigg|_{\vec{p}_{\mathrm{p}} = \vec{p}_\mathrm{i}/{A}}.
\end{align}
The invariant yield of nuclei with mass number $A$, $E_{\mathrm{i}}(\mathrm{d}^{3}N_{\mathrm{i}}/\mathrm{d}p_{\mathrm{i}}^{3})$, is related to the one of protons $E_{\mathrm{p}}(\mathrm{d}^{3}N_{\mathrm{p}}/\mathrm{d}p_{\mathrm{p}}^{3})$, which is expected to be identical to the one of neutrons at midrapidity and LHC energies.
The momentum of the proton $\vec{p}_\mathrm{p}$ is given by the one of the nuclei $\vec{p}_\mathrm{i}$ divided by their mass number.
The coalescence parameter is related to the production probability of the nucleus via this process and can be calculated from the overlap of the nucleus wave function and the phase space distribution of the constituents via the Wigner formalism \cite{Bellini:2018epz}.

\section{Coalescence parameter}

\begin{figure}[hbt]
\centering
\includegraphics[width=0.67\textwidth]{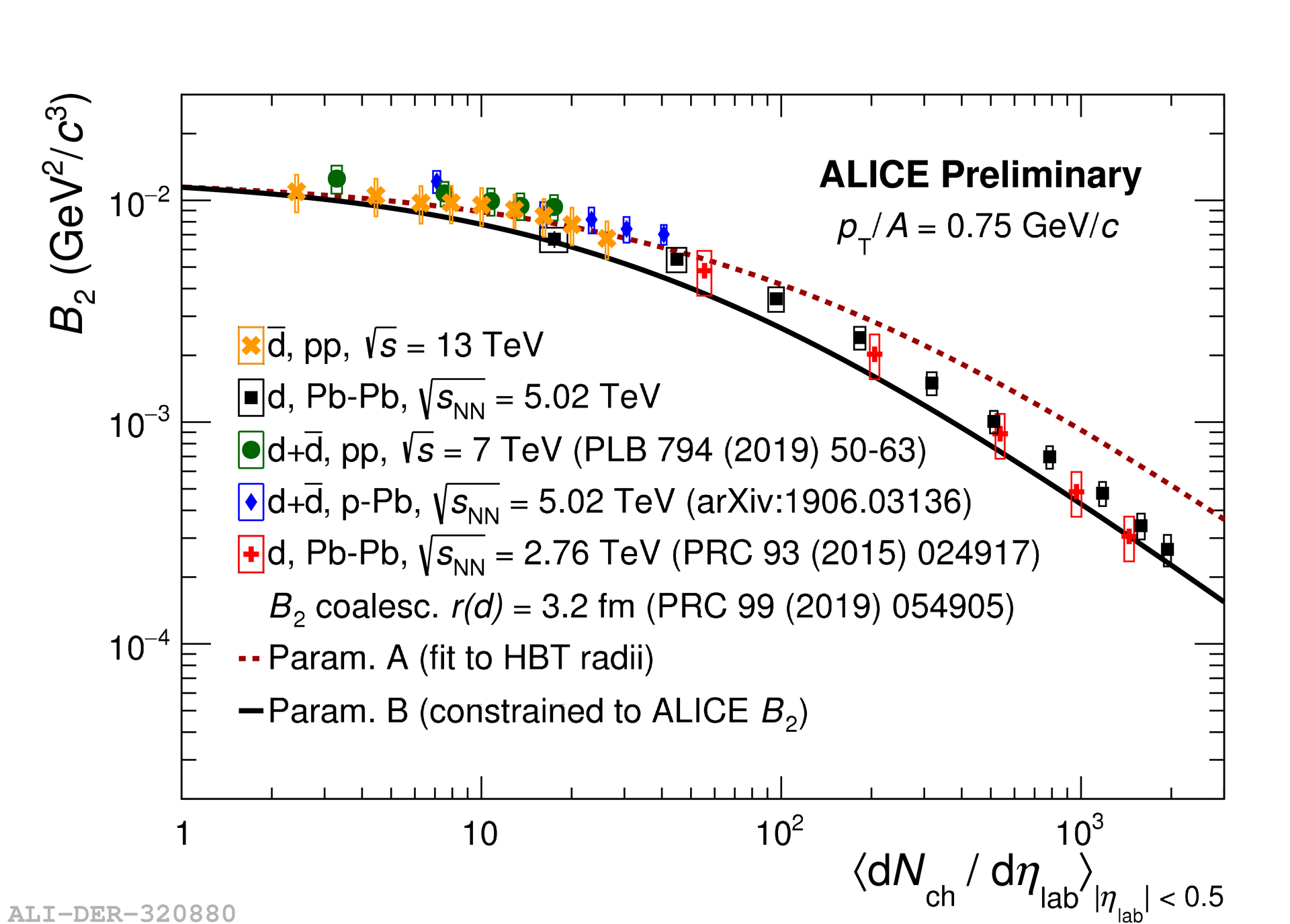}
\caption{$B_2$ as a function of the mean charged-particle multiplicity density $\langle\text{d}N_\text{ch}/\text{d}\eta_\mathrm{lab}\rangle$ for a fixed value of $p_\mathrm{T}/A = 0.75$ GeV/$c$ is compared to the coalescence  calculations from \cite{Bellini:2018epz}. Two different parametrizations for the system size as a function of $\langle\text{d}N_\text{ch}/\text{d}\eta_\mathrm{lab}\rangle$ are used.}
\label{Fig:B2vsMult}
\end{figure}

The evolution of $B_A$ with the system size can be studied by measuring it as a function of $p_\mathrm{T}/A$ for several collision systems and energies \cite{Acharya:2019rgc, Acharya:2019rys, Adam:2015vda}.
For a fixed value of $p_\mathrm{T}/A = 0.75$ GeV/$c$, $B_2$ is shown as a function of the mean charged-particle multiplicity density $\langle\text{d}N_\text{ch}/\text{d}\eta_\mathrm{lab}\rangle$ at midrapidity in figure \ref{Fig:B2vsMult}.
The measurements show a smooth transition from low charged-particle multiplicity densities, which refer to a small system size, to large ones. This indicates that the production mechanism in small systems evolves continuously to the one in large systems and even a single mechanism sensitive to the system size could be possible.
The experimental results are compared with calculations obtained from the coalescence approach \cite{Bellini:2018epz} for two different parameterizations for the system size as a function of $\langle\text{d}N_\text{ch}/\text{d}\eta_\mathrm{lab}\rangle$. The theoretical calculations agree with the trend observed in data.

\section{Ratio of integrated nuclei and proton yields}

\begin{figure}[hbt]
\centering
\includegraphics[width=0.68\textwidth]{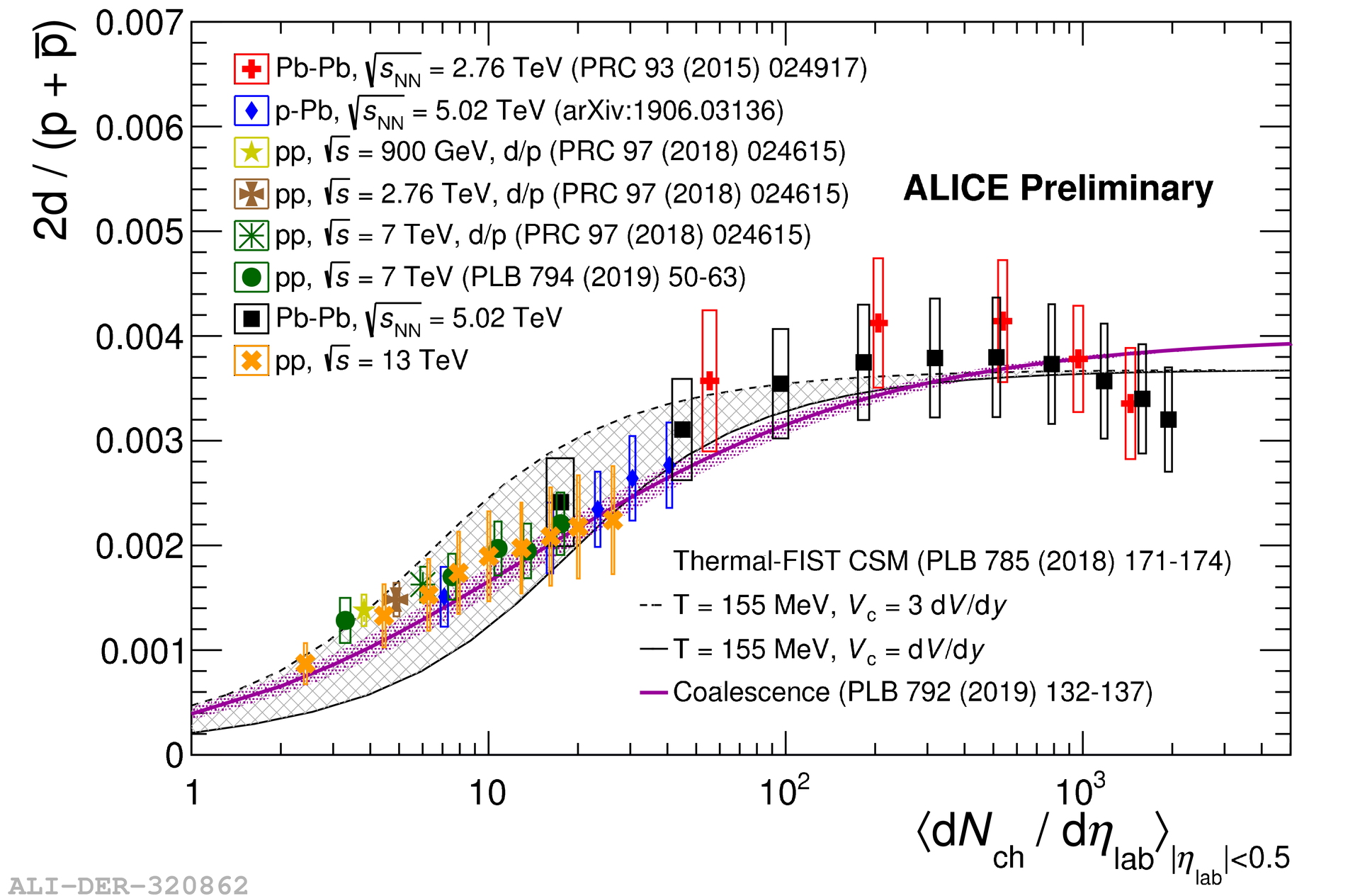}\\
\vspace{0.3cm}
\includegraphics[width=0.65\textwidth]{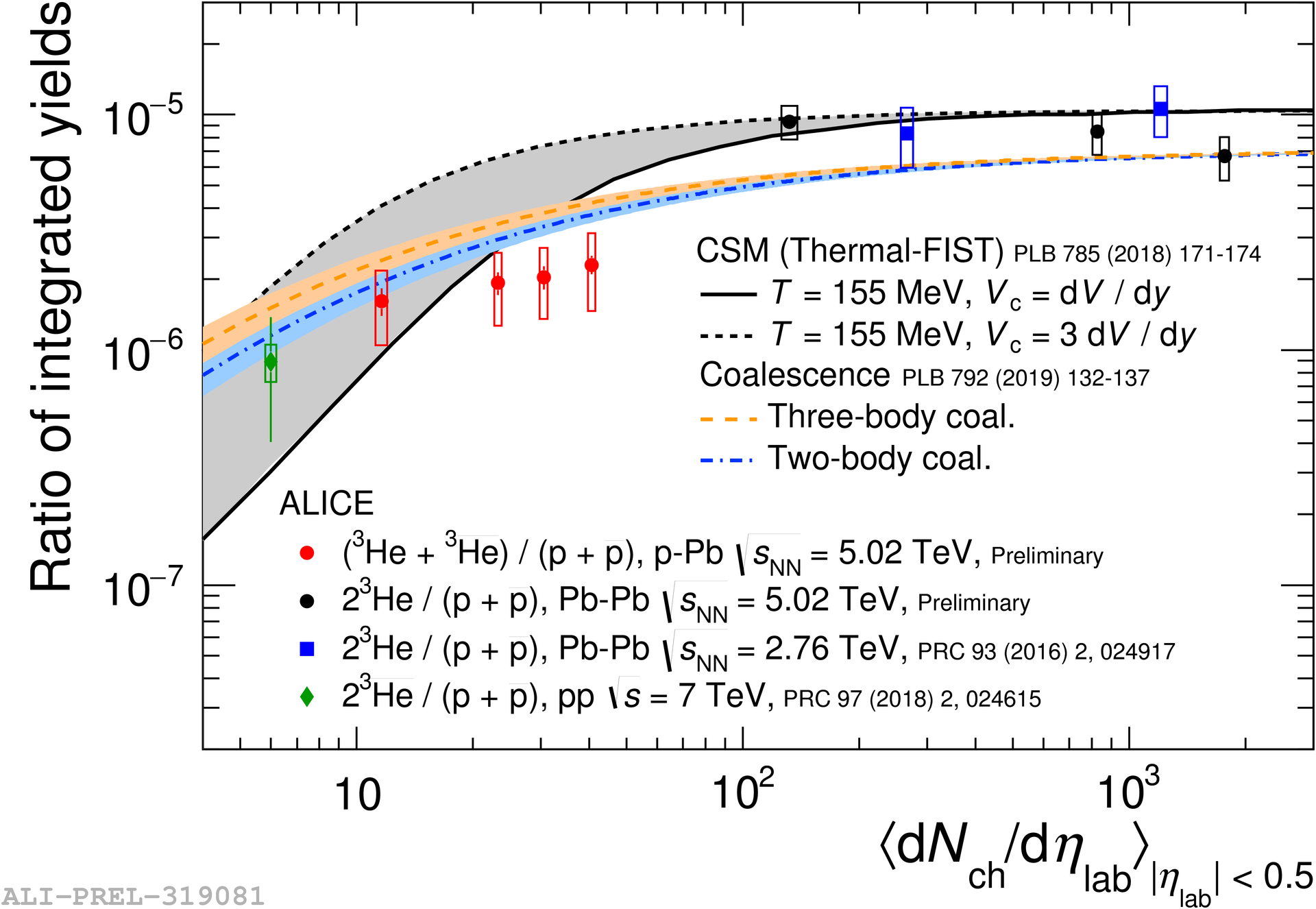}
\caption{The deuteron-to-proton (top panel) and $^3$He-to-proton (bottom panel) ratios in pp, p--Pb and Pb--Pb collisions at different collision energies are shown as a function of the mean charged-particle multiplicity density.
The expectations for the canonical statistical hadronization model, Thermal-FIST \cite{Vovchenko:2018fiy}, and the coalescence approaches \cite{Sun:2018mqq} are shown. For the thermal model, two different values of the correlation volume are displayed.  The uncertainties of the coalescence calculations, which are due to the theoretical uncertainties on the emission source radius, are denoted as shaded bands.}
\label{Fig:RatioToProton}
\end{figure}

Another observable used to learn whether the SHM or the coalescence approach describe the production of nuclei more accurately is the ratio of the integrated yield of the nuclei to that of the protons. It can be studied as a function of $\langle\text{d}N_\text{ch}/\text{d}\eta_\mathrm{lab}\rangle$ as well. A smooth increase of this ratio with the system size is observed, reaching a constant value at mean charged-particle multiplicity densities seen in Pb--Pb collisions.
 Figure \ref{Fig:RatioToProton} shows the deuteron-to-proton and $^3$He-to-proton ratios measured in pp, p--Pb and Pb--Pb collisions \cite{Acharya:2019rgc, Acharya:2017fvb, Acharya:2019rys, Adam:2015vda}.
 The two ratios have a similar trend with $\langle\text{d}N_\text{ch}/\text{d}\eta_\mathrm{lab}\rangle$, however, the increase from the pp to the Pb--Pb results is about a factor of 3 larger for $^3$He/p than for d/p.
The canonical statistical model (CSM) is able to reproduce the trend observed in data thanks to the stronger canonical suppression of the nuclei yields for small system sizes and the saturation towards the grand canonical value at larger values. Especially for the deuteron results at low charged-particle multiplicity, there seems to be a preference towards a larger correlation volume.
It is important to mention that the version of the CSM shown is not able to describe the measured $\phi/\pi$ ratio and struggles with the $p/\pi$ and $K/\pi$ ratios \cite{Vovchenko:2019kes}.
The observed evolution of the ratio is also well described by the coalescence approach due to the increasing phase space. For high charged-particle multiplicity densities, the coalescence calculations seem to underestimate  the ratio of $^3$He-to-protons.

\section{Elliptic and triangular flow}

\begin{figure}[hbt]
\centering
\includegraphics[width=0.65\textwidth]{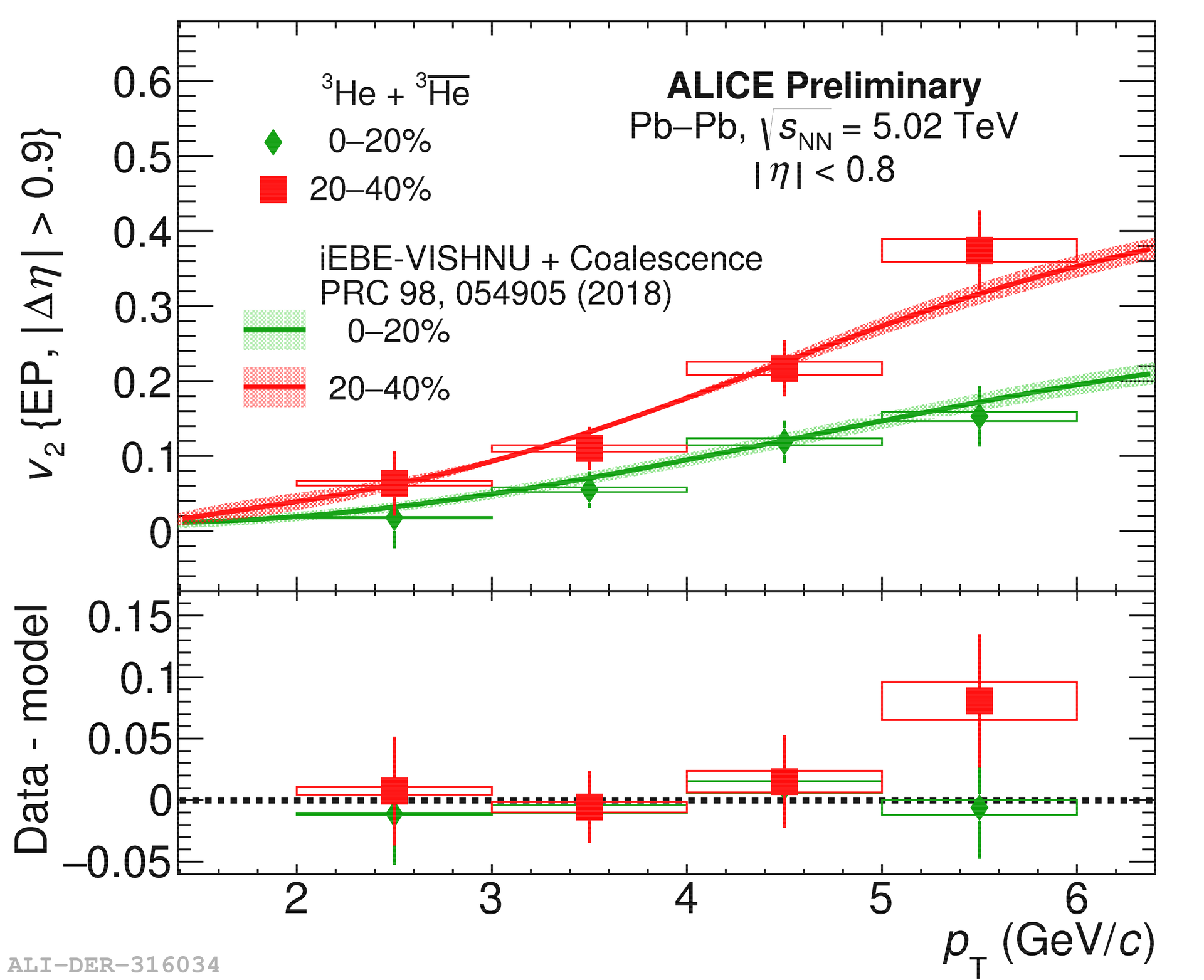}
\caption{Elliptic flow of (anti-)$^{3}$He measured in Pb--Pb collisions at $\sqrt{s_\mathrm{NN}} = 5.02$ TeV in the centrality classes 0--20$\%$ and 20--40$\%$ in comparison with the predictions from a coalescence model based on phase-space distributions of protons and neutrons generated from the iEBE-VISHNU hybrid model with AMPT initial conditions \cite{Zhao:2018lyf}. The model predictions are shown as lines and the bands represent their statistical uncertainties. The differences between data and model are shown in the lower panel for both centrality classes. The statistical uncertainties of the data and the model are added in quadrature. Vertical bars and boxes represent the statistical and systematic uncertainties, respectively.}
\label{Fig:v2Advanced}
\end{figure}

Additional important information about the production process of light (anti-)nuclei can be obtained by studying  their participation in collective flow. New results for the elliptic flow $v_2$ of deuterons and $^3$He, as well as the triangular flow $v_3$ of deuterons, in  Pb--Pb collisions at $\sqrt{s_\mathrm{NN}} = 5.02$ TeV are presented.
The hydrodynamic-inspired Blast-Wave model \cite{Schnedermann:1993ws} and the coalescence approach predictions for the elliptic flow $v_2$ of (anti-)$^3$He are only compatible with the measurements thanks to the large statistical uncertainties (not shown). More state-of-the-art coalescence results, which combine a hydrodynamical simulation (iEBE-VISHNU) with later coalescence of protons and neutrons \cite{Zhao:2018lyf}, provide a good description of the measurement in the two centrality ranges 0--20$\%$ and 20--40$\%$ not only for $v_2$ of $^3$He, as can be seen in figure \ref{Fig:v2Advanced}, but also for the elliptic and triangular flow of deuterons.
The flow of nuclei might challenge theoretical descriptions, thus, a more advanced result for the SHM as well as coalescence results for more peripheral events would be appreciated.

\section{Conclusions}

The measurements of the coalescence parameter as well as the ratio of the integrated yields of nuclei and protons as a function of the charged-particle multiplicity density at midrapidity can be described by the coalescence approach, taking the size of the nucleus and the system into account.
The CSM predicts a similar evolution of the $^3$He-to-proton as well as the d/p ratios with the system size as observed in the measurements. However, it shows discrepancies for other observables like $\phi/\pi$.
The theoretical description of the measured elliptic flow of deuterons and $^3$He in Pb--Pb collisions requires state-of-the-art coalescence calculations while a full SHM result is not available.
A more refined relation between the system size and the charged-particle multiplicity density, as well as future more precise measurements and calculations, will help to challenge the theoretical understanding of the production of (anti-)nuclei in ion collisions in more details.

\bibliographystyle{JHEP}
\bibliography{NucleiBiblio}

\end{document}